\documentclass[aps,print,showpacs,twocolumn]{revtex4}%
\usepackage{amsfonts}
\usepackage{amsmath}
\usepackage{amssymb}
\usepackage{graphicx}%
\begin{document}
\author{Gang Chen$^{a,b\ },$\ J.-Q. Liang$^{a}$}
\affiliation{$^{a}$Institute of Theoretical Physics, Shanxi University, Taiyuan 030006, China}
\affiliation{$^{b}$Department of physics, Shaoxing College of Arts and Sciences, Shaoxing
312000, China}
\title{{\LARGE Quantum Phase Transition in Finite-Size Lipkin-Meshkov-Glick Model }}

\begin{abstract}
Lipkin model of arbitrary particle-number $N$ is studied in terms of exact
differential-operator representation of spin-operators from which we obtain
the low-lying energy spectrum with the instanton method of quantum tunneling.
Our new observation is that the well known quantum phase transition can also
occur in the finite-$N$ model only if $N$ is an odd-number. We furthermore
demonstrate a new type of quantum phase transition characterized by
level-crossing which is induced by the geometric phase interference and is
marvelously periodic with respect to the coupling parameter. Finally the
conventional quantum phase transition is understood intuitively from the
tunneling formulation in the thermodynamic limit.

\end{abstract}

\maketitle

The Lipkin-Meshkov-Glick (LMG) model of many-body two-level system\cite{1,2,3}%
, which was invented originally for the study of nuclear giant
monopole\cite{4}, is exactly solvable\cite{5,6,7} and has wide applications in
the statistical mechanics of mutually interacting spins\cite{8,9} and the
Bose-Einstein condensates\cite{10}. Recently, it has been regarded as an
important model to explore the novel relation between the quantum entanglement
and quantum phase transition (QPT)\cite{11,12,13,14,15}. QPT describing
structural change of the ground-state energy-spectrum of many-body systems
associated with the variation of coupling parameters has attracted
considerable attentions in the modern theoretical and experimental
communities\cite{16}. As a matter of fact QPT originates from singularity of
the energy spectrum\cite{17} and any non-analyticity point in the ground-state
energy of the infinite-lattice-system can be identified as a QPT. There can be
a level-crossing where the excited-state and ground-state levels interchange
at the critical-point, creating non-analyticity of the ground-state energy
which is a function of the coupling parameter\cite{16}. It has been
demonstrated that the LMG model described by the Hamiltonian\cite{18}
\begin{equation}
\widehat{H}=-\lambda\sum_{i<j}(\widehat{\sigma}_{x}^{i}\widehat{\sigma}%
_{x}^{j}\pm\gamma\widehat{\sigma}_{y}^{i}\widehat{\sigma}_{y}^{j})-h\sum
_{i}\widehat{\sigma}_{z}^{i}, \label{1}%
\end{equation}
undergoes a QPT from a deformed to a normal phases at the critical-value of
coupling parameter $h_{c}=\lambda^{\prime}=N\lambda$, where the parameter
$\lambda^{\prime}$ is used to avoid the formal infinity\cite{18,20} in the
thermodynamic limit $N\rightarrow\infty$ and $\widehat{\sigma}_{\alpha}$ is
the Pauli matrices. In the deformed phase when $h\leq h_{c}$ the ground state
is degenerate while is non-degenerate in the normal phase when $h\geq h_{c}$.
For our convenience we confine the parameters $\lambda>0$ , $0<\gamma<1$ and
consider the two cases denoted by $"\pm"$ signs separately. In spin language
the deformed-phase corresponds to a long-range magnetic order and may be
regarded as a ferromagnet, while the transverse-magnetic-field $h$ lowering
the potential-barrier induces quantum tunneling between two generate-states
and eventually destroy the magnetic-order at the critical point $h=h_{c}$. A
similar QPT was indeed observed with the ferromagnetic moment vanishing
continuously at a quantum critical point of the transverse field in the
low-lying magnetic excitations of the insulator LiHoFe$_{4}$ \cite{19}.

The study of scaling properties at the critical point for a finite-$N$ model
is of primer interest and\textbf{\ }has become a hard topic
recently\cite{18,20}. Based on mean-field approximation it is found that
particle-number-dependence of the energy gap between the ground and
first-excited states behaviors as $\exp[-N]$ in the deformed phase\cite{18,20}%
. However QPT \ from deformed to normal phases, which we called the
conventional QPT (CQPT), actually does not exist for a finite-$N$ \cite{8}
since the quantum tunneling can remove the degeneracy leading to the tunnel
splitting $\Delta E$ even at $h=0$ and thus the ground-state energy-structure,
strictly speaking, does not have macroscopic change. The energy gap $\Delta
E$\ vanishes in the thermodynamic limit $N\rightarrow\infty$ and the ground
state becomes degenerate. The quantum tunneling effect of the finite-$N$
model, which gives rise to a rich structure of low-lying energy spectrum has
not yet been studied explicitly. Based on the tunneling effect obtained in
this Letter we demonstrate that the CQPT can occur in the finite-$N$ model if
$N$ is an odd-number due to the quenching of quantum tunneling induced by
quantum-phase interference and the magnetic-order survives from the tunneling
till the potential-barrier separating the two degenerate-states is suppressed
sufficiently low by the transverse field.

Moreover we find a new type of QPT characterized by the level-crossing which
is induced also by quantum-phase-interference. The ground and first-excited
states interchange alternatively with the increase of $h$ resulting in
periodic QPT. The finite-$N$ Lipkin model is studied in terms of the exact
differential-operator representation of the spin-operators which gives rise to
a potential-barrier tunneling Hamiltonian and the low-lying energy spectrum is
obtained by means of instanton method. From the viewpoint of tunneling the
CQPT can be understood intuitively since the potential-barrier height
decreases with the increase of $h$ and finally disappears at the critical
field-value. The level-crossing, however, does not appear at all in the CQPT.

We adopt the giant-spin operators defined by $\widehat{S}_{l}=\sum_{i}%
^{N}\widehat{\sigma}_{l}^{i}/2$ ($l=x,y,z$) to convert the $N$-particle model
into an anisotropic giant-spin Hamiltonian
\begin{equation}
\widehat{H}=-\lambda(\hat{S}_{x}^{2}\pm\gamma\hat{S}_{y}^{2})-h\hat{S}_{z}.
\label{2}%
\end{equation}
with the total spin-quantum-number\textbf{ }$s=N/2$, \textbf{\ }$N/2-1$,
\textbf{\ }$N/2-2\cdot\cdot\cdot,1/2$, or $0$ depending on odd- or even-$N$
and a factor $2$ absorbed into the parameters $\lambda$ and $h$\textbf{.}
Because of the ferromagnetic interaction between spins, $\lambda>0$, the
ground state and the first-excited states always lie in the subspace of
maximum spin \cite{18} $s=N/2$. The Hamiltonian of $"+"$ sign describes a
giant-spin with an easy axis-x (lowest energy) and a hard axis-z (highest
energy). The external magnetic field $h$ is applied along the hard axis, which
is critical in generating the QPT of level-crossing. For the $"-"$ sign case
y-axis becomes the hard-axis and the magnetic field is then along the medium axis.

We start from stationary Schr\"{o}dinger equation $\widehat{H}\Phi
(\varphi)=E\Phi(\varphi)$ with Hamiltonian of $"-"$ sign and a
coordinate-frame rotation such that the hard-axis and easy-axis are along z
and x respectively for the sake of convenience, where $\varphi$ can be
visualized as the azimuthal angle of the giant-spin vector and the generating
function\cite{21} $\Phi(\varphi)$ can be constructed in terms of the
conventional\ eigenstates of $\widehat{S}_{z}$ such as $\Phi(\varphi
)=\sum_{m=-s}^{s}C_{m}\exp(im\varphi)/\sqrt{(s-m)!(s+m)!}$, which obviously
satisfies the boundary condition, $\Phi(\varphi+2\pi)=\exp(2\pi is)\Phi
(\varphi)$. Thus, we have periodic wave functions for even-$N$ and
anti-periodic wave functions for odd-$N$. The explicit form of the spin
operator acting on the wave function $\Phi(\varphi)$ is seen to be\cite{21}
$\widehat{S}_{x}=s\cos\varphi-\sin\varphi\frac{d}{d\varphi},\quad\widehat
{S}_{y}=s\sin\varphi+\cos\varphi\frac{d}{d\varphi},$ and $\widehat{S}%
_{z}=-i\frac{d}{d\varphi}$. Substituting these differential operators into the
Schr\"{o}dinger equation we obtain $\{-\alpha(1-\delta\sin^{2}\varphi
)\frac{d^{2}}{d\varphi^{2}}-[\lambda(s-\frac{1}{2})\sin2\varphi-h\sin
\varphi]\frac{d}{d\varphi}+[\lambda(s^{2}\cos^{2}\varphi+s\sin^{2}%
\varphi)-hs\cos\varphi]\}\Phi(\varphi)=E\Phi(\varphi)$, where $\alpha$
$=\lambda(1+\gamma)$, $\delta=1/(1+\gamma)$. In the new variable $x$ defined
as $x(\varphi)=\int_{0}^{\varphi}d\varphi^{\prime}/\sqrt{1-\delta\sin
^{2}\varphi^{\prime}}=F(\varphi,\delta)$, which is the incomplete elliptic
integral of the first kind with modulus $\sqrt{\delta}$, the trigonometric
functions $\sin\varphi$ and $\cos\varphi$ become the Jacobian elliptic
functions $sn(x)$ and $cn(x)$ with the same modulus, respectively\cite{22}.
Using the transformation such that $\Phi\lbrack\phi(x)]=dn^{s}(x)\exp
\{-h\tan^{-1}[cn(x)/\sqrt{\gamma}]/2\lambda\sqrt{\gamma}\}\psi(x)$, where
$dn(x)=\sqrt{1-\delta sn^{2}(x)}$ is also a Jacobian elliptic function, the
stationary Schr\"{o}dinger equation becomes%

\begin{equation}
\lbrack-\alpha\frac{d^{2}}{dx^{2}}+V(x)]\psi(x)=E\psi(x), \label{3}%
\end{equation}
with the scalar potential given by $V(x)=[\zeta cn^{2}(x)-h(s+\frac{1}%
{2})cn(x)+h^{2}/4\alpha]/dn^{2}(x)$, \quad$\zeta=\lambda s(s+1)-h^{2}/4\alpha
$. The boundary condition of wave function becomes $\psi(x+4K)=\exp(2\pi
is)\psi(x)$ with $K$ denoting the complete elliptic integral of the first
kind. In the absence of external field $h=0$ the scalar potential $V(x)$ is
symmetric with respect to the coordinate origin (see Fig.(1), red-line) with
potential minima located at $x_{\pm}=$ $\pm K$ in one-period corresponding to
the original azimuthal angles $\varphi_{\pm}=0,\pi$ which are two equilibrium
orientations of the giant-spin. Disregarding the negligibly small quantum
tunneling the two ground states are degenerate. The external field suppresses
the potential barrier and thus increases the tunneling rate which destroys the
magnetic order gradually. We now calculate the tunneling rate between the two
degenerate-states. To this end we begin with the tunneling induced Feynman
propagator through the potential barrier%
\begin{equation}
<x_{-},\beta|x_{+},-\beta>=<x_{-}|e^{-2\beta\widehat{H}}|x_{+}>=\int
\mathcal{D\{}x\}e^{-S} \label{4}%
\end{equation}
where $S=\lim_{\beta\rightarrow\infty}\int_{-\beta}^{\beta}\mathcal{L}%
_{e}d\tau$ is the Euclidean action evaluated along the tunneling trajectory of
a pseudoparticle in the barrier region called instanton with the Euclidean
Lagrangian $\mathcal{L}_{e}=\dot{x}^{2}/4\alpha+U(x)$. $\dot{x}=dx/d\tau$
denotes the imaginary-time derivative with $\tau=it$. The quantum tunneling
removes the degeneracy of ground-states $\left\vert \psi_{c\pm}\right\rangle $
in two potential wells located at $x_{\pm}$ respectively which describe the
two equilibrium-orientations of the giant-spin and thus may be called the
Schr\"{o}dinger cat states. In the two-level approximation we have
$\widehat{H}\left\vert 1\right\rangle =E_{1}\left\vert 1\right\rangle $,
$\widehat{H}\left\vert 0\right\rangle =E_{0}\left\vert 0\right\rangle $, with
$\left\vert 0\right\rangle $, $\left\vert 1\right\rangle =(\left\vert
\psi_{c+}\right\rangle \pm\left\vert \psi_{c-}\right\rangle )/\sqrt{2}$ being
the ground and first-excited states. The low-lying energy spectrum can be
evaluated as $E_{1,0}=\overline{E}\pm\Delta E/2$ with $\overline{E}%
=<\psi_{c\pm}|\widehat{H}|\psi_{c\pm}>$, and $\Delta E=E_{1}-E_{0}%
=-(<\psi_{c+}|\widehat{H}|\psi_{c-}>+<\psi_{c-}|\widehat{H}|\psi_{c+}>)$.
Inserting the complete set $\left\vert \psi_{c\pm}\right\rangle $ in the
transition amplitude eq.(4) the tunnel splitting can be derived from the
path-integral
\begin{equation}
\Delta E\sim\frac{e^{2\beta\overline{E}}}{2\beta\psi_{c+}(x_{+})\psi
_{c-}^{\ast}(x_{-})}\int\mathcal{D\{}x\}e^{-S},
\end{equation}
which can be evaluated in terms of the stationary-phase approximation that
$\int\mathcal{D\{}x\}e^{-S}\sim Ie^{-S_{c}}$ where $S_{c}$ is the action along
the classical trajectory of instanton $x_{c}(\tau)$ i.e. the solution of
classical equation of motion $\delta S=0$. $I=\int\mathcal{D\{}\eta
\}e^{-\frac{\delta^{2}S}{\delta x\delta x}|_{x_{c}}}$ denotes the contribution
of quantum fluctuation around the classical trajectory such that
$x(\tau)=x_{c}(\tau)+\eta(\tau)$ with $\eta$ being the small fluctuation.
Working out the path-integral with both the clockwise and anti-clockwise under
barrier rotations (see Fig(1)) and wave functions $\psi_{c\pm}(x_{\pm})$
noticing the boundary condition we obtain the level splitting%
\begin{equation}
\Delta E=Pe^{-W}\sqrt{\cosh\chi+\cos2\pi s}%
\end{equation}
with $P=2^{7/2}\{U_{0}^{3}\lambda(1+\gamma)(1-\varrho^{2})^{5}/[\gamma
/(1+\gamma)+\varrho^{2}/(1+\gamma)]^{2}\pi^{2}\}^{1/4}$ , $W=\sqrt
{U_{0}(1+\gamma)/\lambda}\ln[(1+\sqrt{(1-\varrho^{2})/(1+\gamma)}%
)/(1-\sqrt{(1-\varrho^{2})/(1+\gamma)})]$, and $\chi=2\varrho\sqrt
{U_{0}/\lambda\gamma}\tan^{-1}[\sqrt{\gamma(1-\varrho^{2})/(1+\gamma)}%
/\varrho]$, where $U_{0}=[\zeta_{1}-\zeta_{3}/\gamma+\sqrt{(\zeta_{1}%
-\zeta_{3}/\gamma)^{2}+\zeta_{2}^{2}/\gamma}]/2$ and $\varrho=\zeta_{2}%
/2U_{0}$ with $\zeta_{1}=\lambda s(s+1)-h^{2}/4\lambda(1+\gamma)$, $\zeta
_{2}=h(s+1/2)$ and $\zeta_{3}=h^{2}/4\lambda(1+\gamma)$. When $h=0$ we have
$\chi=0$ \ and the square-root factor in eq.(6), which is a direct result of
quantum phase interference of wave functions in eq.(5) between two tunnel
paths (Fig.(1)), is $\sqrt{2}$ for even-$N$. The quantum tunneling destroys
the magnetic order and the ground state $\left\vert 0\right\rangle $ becomes
paramagnet. There is no QPT for finite-$N$. However the situation is different
for the odd-$N$ case where the square-root factor in eq.(6) becomes zero
leading to vanishing tunnel splitting. The degeneracy of ground states at
$h=0$ cannot be removed by tunneling similar to the Kramer's degeneracy in
spin system and the ferromagnetic ground states $\left\vert \psi_{c\pm
}\right\rangle $ remain. Increase of the magnetic field suppresses the
potential barrier seen from Fig.(1) and enhances the tunneling rate which
destroys the magnetic order gradually. The system thus undergoes a smooth
second order QPT. The critical transition point $h_{c}=\lambda(1-\gamma
)(s+1/2)+\lambda\sqrt{(1-\gamma)^{2}(s+1/2)^{2}+4\gamma s(s+1)}$ can be
determined precisely from the condition that the central barrier-height
vanishes and the double-well of the potential becomes a single one (see
Fig(1)), which reduces exactly to $h_{c}^{th}=\lambda^{\prime}$ in the
thermodynamic limit\cite{18,20}. The numerical simulation of the low-lying
levels is shown in Fig.(2) the energy gap obtained from which agrees with our
gap formula eq.(6) perfectly for any $s$. From eq.(6) the finite-size scaling
behavior of the energy gap is $\Delta E\sim\exp[-N]$ which vanishes in the
thermodynamic limit no matter $N$ is even or odd.

We now consider the most interesting case of $"+"$ sign where the external
field is along the hard-axis (z-axis). The Hamiltonian operator does not
possess a satisfactory differential-operator representation\cite{23}. The
Feynman propagator eq.(4) in angle variables may be evaluated with the
spin-coherent-state path-integrals with the spin coherent state defined by
$\widehat{S}\mathbf{\cdot n}|\mathbf{n}>=s|\mathbf{n}>$ where $\mathbf{n=(}%
\sin\theta\cos\varphi,\sin\theta\sin\varphi,\cos\theta)$ is a unit
vector\cite{24,25}. Regarding $\varphi$, $p=s\cos\theta$ as canonical
variables and integrating over variable $p$ we have the effective Lagrangian
given by $\mathcal{L}=\frac{1}{2}m(\varphi)\dot{\varphi}^{2}+A(\varphi
)\dot{\varphi}-V(\varphi),$where $m(\varphi)=1/2\lambda(1-(1-\gamma)\sin
^{2}\varphi)$, $A(\varphi)=s[1-h/2\lambda s(1-(1-\gamma)\sin^{2}\varphi)],$and
$V(\varphi)=\lambda(1-\gamma)s(s+1)\sin^{2}\varphi-h^{2}(1-\gamma)\sin
^{2}\varphi/4\lambda(1-(1-\gamma)\sin^{2}\varphi)$. The periodic potential
$V(\varphi)$ has degenerate vacua located at $\varphi=0$, $\pi$ corresponding
to the equilibrium orientations of the giant spin along $\pm$x directions. We
again consider the tunneling induced propagator between two degenerate vacua.
The total-time-derivative term $A(\varphi)\dot{\varphi}$ coming from the
geometric phase of the path-integral does not affect the classical equation of
motion but gives rise to the quantum phase interference between the clockwise
and anti-clockwise tunneling paths. With the same procedure the tunnel
splitting is found as%
\begin{equation}
\Delta E=Pe^{-W}\cos[\pi(s-\theta_{h})]
\end{equation}
\ with $P=2^{4}\sqrt{s^{3}\lambda^{2}(1-\gamma)^{3/2}(1-\eta)^{5/2}/\pi
(\gamma-\eta)}$,

$W=\sqrt{s(s+1)}\{\ln[(\sqrt{1-\eta}+\sqrt{1-\gamma})/(\sqrt{1-\eta}%
-\sqrt{1-\gamma})]-\sqrt{\eta/\gamma}\ln[(1+\sqrt{\varsigma})/(1-\sqrt
{\varsigma})]\}$ and $\theta_{h}=h/2\lambda\sqrt{\gamma}$, where $\eta
=h^{2}/4\lambda^{2}s(s+1)$ and $\varsigma=(1-\gamma)\eta/\gamma(1-\eta)$. We
now come to the most interesting observation of the paper that the energy gap
is a periodic function of $h$ and the two states interchange periodically. The
the QPT of level-crossing arises at the critical points $h_{c}(n)=2\lambda
n\sqrt{\gamma}$ \ for odd-$N$, \ ($2n+1)\lambda\sqrt{\gamma}$ for even-$N$,
where $n$ is integer. Fig.(3) shows the numerical result of the first-excited
(red-line) and ground (blue-line) energy levels for odd-$N$ (a) and even-$N$
(b) respectively. The absolute value of numerical energy-gap as a function of
$h$ agrees qualitatively with eq.(7) and the agreement becomes perfect for
large $N$ (see Fig(3)-insert) since the spin-coherent-state-path-integral
method is valid for large $s$. For the odd-$N$ case the ground state at $h=0$
($\theta_{h}=0$) is degenerate since $\cos(\pi s)=0$. When $h\leq h_{c}(1)$
quantum tunneling destroys the degeneracy and ground state becomes quantum
paramagnetic while the first excited state is ferromagnetic. This can be
explained with the Schr\"{o}dinger cat states that $\left\vert \psi_{c\pm
}\right\rangle =\left\vert \mathbf{n}_{\pm}\right\rangle $ where $\widehat
{S}_{x}|\mathbf{n}_{\pm}>=\pm s|\mathbf{n}_{\pm}>$ with the unit vector of
spin-coherent-state along the x-axis. We have $<0|\widehat{S}_{x}|0>=0$, and
$<1|\widehat{S}_{x}|1>=s$. Increase of the coupling $h$ crossing over the
first critical point $h_{c}(1)$ the first-excited and ground states
interchange and system becomes the ferromagnetic phase. The interchange takes
place again at the second critical point and is particularly useful in the
quantum computing since the two phases can be used to realize a qubit in some
practical spin-system.\textbf{ }The magnetization as a function of the
magnetic field can be evaluated as $M_{0,1}=-<0,1|\partial\widehat{H}/\partial
h|0,1>=-\partial E_{0,1}/\partial h$. The numerical simulation of the
magnetization is shown in Fig.4. The critical behavior of the energy gap
$\Delta E$ at the critical point $h_{c}$ is found as $\Delta E(h\rightarrow
h_{c})=(\pi/\lambda\sqrt{\gamma})P(h_{c})e^{-W(h_{c})}\left\vert
h-h_{c}\right\vert .$ In the thermodynamic limit the tunnel splitting vanishes
as $\lim_{N\rightarrow\infty}\Delta E\sim\exp[-N]\rightarrow0$ and the QPT of
level-crossing disappears$.$A critical value $h_{c}^{th}=\lambda^{\prime}%
\sqrt{\gamma}$ of CQPT can be determined again from the vanishing
barrier-height, which differs from mean-field value by a factor $\sqrt{\gamma
}$.

In conclusion: with the explicit calculation of tunneling effect we show that
QPT can occur in finite-size LMG model. The QPT of level-crossing discovered
in this paper may provide a possible technique to realize qubit in quantum
computer and a paper explaining this procedure will be published elsewhere.

This work was supported by the Natural Science Foundation of China under Grant
No.10475053 and by the Natural Science Foundation of Zhejiang Province under
Grant No.Y605037.

\end{document}